\documentclass{article}
\usepackage[utf8]{inputenc}
\usepackage{hyperref}
\usepackage{amsmath}
\usepackage{amsfonts}
\usepackage{amssymb}
\usepackage{graphicx}

\title{Cyber-sensorium: An Extension of the Cyber Public Health Framework}
\author{
    Robin Coupland\thanks{Visiting Lecturer, University of Lucerne} \and
    Nathan Taback\thanks{Professor Teaching Stream, Department of Statistical Sciences, University of Toronto}
}
\date{June 9, 2024}

\begin{document}

\maketitle

\begin{abstract}
In response to increasingly sophisticated cyberattacks, a health-based approach is being used to define and assess their impact. Two significant cybersecurity workshops have fostered this perspective, aiming to standardize the understanding of cyber harm. Experts at these workshops agreed on a public health-like framework to analyze cyber threats focusing on the perpetrators’ intent, the means available to them, and the vulnerability of targets. We contribute to this dialogue with the ``cyber-sensorium" concept, drawing parallels between the digital network and a biological nervous system essential to human welfare. Cyberattacks on this system present serious global risks, underlining the need for its protection.
\end{abstract}

\section{Introduction}
The damage caused by and the increasing sophistication of cyberattacks is the stuff of everyday news. It is clear that with the rapid advances in artificial intelligence, such attacks become increasingly easy to perpetrate and more sophisticated in terms of their impact.

The first step in prevention of a problem in any sphere is to gather pertinent data about it. If you want to manage it, measure it! What gets counted gets done! Two organisations are making headway in gathering data about cyberattacks and their impact. Both invoke a notion of “health” \footnote{The WHO Definition of health is: “Health is a state of complete physical, mental, and social well-being and not merely the absence of disease or infirmity.”} in their approach.

\section{CyberPeace Workshop}
The CyberPeace Institute in Geneva held an Expert Meeting workshop \cite{cpi_workshop} in November 2023 with representatives from States, academia, International Organizations, and other civil society actors with the aim of building consensus on a typology of “harms” (or impact) of cyberattacks and – as part of its Cyber Peace Watch program and with a view to developing preventive policies and accountability – a methodology by which data pertaining to such harms might be gathered. A number of important concepts were discussed at the workshop and agreed upon.

First, because of the variety of terms used and through an imperative given by the scientific nature of the initiative, there is a need for definitions. What constitutes a cyberattack? What constitutes “harm”? How and why should a distinction be made between harms that are intended and unintended, foreseen and unforeseen, or tangible and intangible?

Second, the physical or virtual site of harms might be limited to computers, computer systems, and the data contained therein; or there may be knock-on harms to institutions or services and then to people’s health and well-being or the environment. In other words, the full extent of the harm produced by a cyberattack – if known – must be recorded in a systematic manner.

The third concept agreed upon was that an impact-orientated analytical framework given by a theory of violence \cite{coupland_theory} provides a valid parallel for consideration of cyberattacks. The theory posits that any given quantifiable effect of an act of violence in any context can be considered in terms of its impact on the health of the victim(s). This then forces consideration of the determinants (or – in public health terms – “risk factors”) of that impact and how they interact. These determinants always comprise: the intent of the perpetrator(s); their physical capacity given by the weapon(s) to cause the effect; and the vulnerability of the victim(s). Preventive measures always relate to one or more of these determinants.

In terms of cyberattacks, the workshop agreed that determinants of the harms are: the purposeful use by the perpetrator of their technical capacity to perpetrate the attack, and the vulnerability of the systems or people potentially impacted. Importantly, this approach does not involve making a moral or legal judgment about a given attack; the facts are considered in a totally objective perspective. As a result of this discussion, cyberviolence was defined as: “The purposeful use, threat of use, negligent use, or autonomous action of digital and information technologies that directly, indirectly, temporarily or permanently causes either immediate or long-term harm determined as a negative impact on people’s health, their physical security, their economic security, or on the environment.”

The fourth concept - which generates an ambitious undertaking – is that compilation of details about real cyberattacks and the resulting “harms” in a structured database with subsequent analysis of the resulting dataset to indicate accountability and to reveal the where, when, how and what of appropriate preventive measures. A further workshop to advance this undertaking is expected in 2024.

\section{CyberGreen Workshop}
CyberGreen based in New York is also at the forefront of integrating health concepts into cybersecurity, hosting a workshop that attracted technical experts from leading IT entities \cite{cybergreen_report}. CyberGreen’s technical reports advocate for adopting public health strategies in cybersecurity, including the Cyber Belief Model and vital statistics to systematically improve cyber health. This approach aims to transform public health practices into effective cybersecurity measures, enhancing the resilience of digital ecosystems against cyber threats. This framework, detailed on Adam Shostack's website \cite{shostack}, serves as a blueprint for understanding and combating cyber risks through a health-centric lens.

A keynote by one of the authors, Nathan Taback \cite{Taback_keynote}, delved into public health paradigms applied to cyber threats. He emphasized the necessity of a systemic approach akin to monitoring the well-being of a biological organism to safeguard the digital ecosystem’s integrity. This aligned with the workshop's agenda, which scrutinized the global health of computer systems and the broader internet infrastructure, questioning how cyberattacks affect this "health" beyond direct human impact. Taback suggested that data-driven and public health-oriented strategies could enhance our understanding and mitigation of cyber risks. Clear metrics and the communication of cyber threats without inducing “alarm fatigue” \cite{alarm_fatigue} underlines a nuanced approach to cybersecurity as a collective health issue. Hence, while CyberGreen and the CyberPeace Institute may adopt different strategies, both resonate with the broader objective of conceptualizing cyberattacks within a health-oriented framework, advocating for a robust analytical and sensitively communicated defense mechanism against the digital age's pandemics.

\section{Cyber-sensorium}
We propose an extension to thinking about a notion of health in relation to cyberattacks. When considered together, all computers, computer services, means by which the technology connects, resulting data banks, and the ‘internet of things’ carry many parallels to an animal’s brain, spinal cord, and peripheral nerves. This parallel becomes more pertinent with the advances in and widespread application of generative artificial intelligence. For want of a better term, we propose that this phenomenon be termed the “cyber-sensorium.”

Whether or not the reader agrees with this analysis, the undeniable fact is that the cyber-sensorium dominates every aspect of our lives. It interfaces with and fuels all the emergent and essential features of human existence such as innovation, law, trade, security, and health care. In brief, the cyber-sensorium is now an integral part of humanity’s well-being and its integrity can only take on increasing importance for us. This underscores how important a notion of health is in relation to the cyber-sensorium. At a global level, the health of the cyber-sensorium determines our well-being, and so attacks on it are an attack on humanity.

Is a notion of neurological health (including mental health) pertinent to the cyber-sensorium? The cyber-sensorium can, like a nervous system, be considered in terms of structure and function. The brain can suffer physical damage from, for example, trauma or a vascular accident. Examples of brain dysfunction with normal structure are depression and schizophrenia. People’s mental well-being can also be impacted by the brain’s normal function in the face of abnormal circumstances such as stress or grief. The spinal cord and peripheral nerves likewise can also be damaged by trauma, compression, or disease. The cyber-sensorium can suffer structural damage through physical attacks on, for example, computers, databanks, or cables. Dysfunction with normal structure can result through breaches of confidentiality, integrity, and accessibility of data. Examples of negative impact through normal structure and function are phishing, insertion of malware, and misinformation campaigns. Furthermore, harms within the cyber-sensorium can manifest themselves without; examples are the hemi-paralysis after a stroke or the typical problems of locomotion associated with Parkinson’s disease.

Therefore, the definition of cyberviolence as given by the Cyber Peace Institute could be expanded to include harms to the cyber-sensorium as follows: “The purposeful use, threat of use, negligent use, or autonomous action of digital and information technologies that directly, indirectly, temporarily or permanently causes either immediate or long-term harm determined as dysfunction of any part of the cyber-sensorium, or a negative impact on people’s health, their physical security, their economic security, or on the environment.”

\section{Conclusion}
Is considering a cyberattack in terms of an impact on the “health” of the cyber-sensorium valid? If so, the theory of violence is pertinent and therefore the impact of a cyberattack on any part of the cyber-sensorium results from the purposeful use by the perpetrator of their technical capacity to perpetrate the attack, and the vulnerability of that part of the cyber sensorium in question. This could provide the basis for an impact-orientated analytical framework applicable to cyberattacks. Further, this approach might help to make sense of the bigger picture of humanity’s interaction with the cyber-sensorium we have created and why this interaction must be safeguarded. The implications for not caring for the cyber-sensorium are considerable. Our prosperity, our well-being, our security, and the environment are all at stake.

\bibliographystyle{unsrt}
\bibliography{references.bib}

\end{document}